\documentclass{raa}          

\usepackage{graphicx,times}             %for PS/EPS graphics inclusion, new
\usepackage{natbib}
\usepackage{amssymb,amsmath}
\bibpunct{(}{)}{;}{a}{}{,}

\def\msol {\hbox{M$_{\odot}$}}

\def\kmss    {km\, s$^{-1}$}
\def\kms     {km\, s$^{-1}$\,}
\def\Rn2 {\Romannum{2} }
\def\Rn1 {\Romannum{1} }

\def\apj {ApJ}
\def\aap {A\&A}
\def\aaps {A\&AS}
\def\apjl {ApJL}
\def\apjs {ApJS}
\def\mnras {MNRAS}
\def\aj {AJ}
\def\nat {Nature}

\begin{document}

   \title{Fallback in bipolar Planetary Nebulae ?}
%\footnotetext{$*$ Supported by the National Natural Science Foundation of China.}
%}
%   \subtitle{I. Place Your Subtitle Here}

   \volnopage{Vol.0 (20xx) No.0, 000--000}      %%preserved for Editor. DOn't remove!
   \setcounter{page}{1}          %%starting page, preserved for Editor. DOn't remove!

   \author{Willem A.~Baan
      \inst{1,2,3}
    \and Hiroshi Imai
      \inst{2,4}
      \and Gabor Orosz
      \inst{2,3,5}
   }

   \institute{Netherlands Institute for Radio Astronomy ASTRON, 79901 PD Dwingeloo, the Netherlands, {baan@astron.nl}\\
        \and
Center for General Education, Institute for Comprehensive Education, Kagoshima University, 1-21-30 Korimoto, 
Kagoshima 890-0065, Japan\\
        \and
Xinjiang Astronomical Observatory, 150 Science 1-Street, Urumqi, Xinjiang 830011, P.R. China\\
        \and
Amanogawa Galaxy Astronomy Research Center, Graduate School of Science and Engineering, Kagoshima University, 1-21-35 Korimoto, Kagoshima 890-0065, Japan\\
 \and
School of Natural Sciences, University of Tasmania, Private Bag 37, Hobart, Tasmania 7001, Australia\\
 \vs\no
   {\small Received~~20xx month day; accepted~~20xx~~month day}}

\abstract{ The subclass of bipolar Planetary Nebulae (PN) exhibits well-defined low-power outflows and 
some show shock-related equatorial spiderweb structures and hourglass structures surrounding these outflows. 
These structures are distinctly different from the phenomena associated with spherical and elliptical PN and 
suggest a non-standard way to simultaneously energise both kind of structures.
This paper presents evidence from the published literature on bipolar PN Hb\,12 and other sources in support of an 
alternative scenario for energising these structures by means of accretion from material shells deposited during 
earlier post-AGB and pre-PN evolutionary stages.
In addition to energising the bipolar outflow, a sub-Eddington accretion scenario could hydrodynamically explain the
spiderweb and outer hourglass structures as oblique shockwaves for guiding the accreting material into the equatorial region 
of the source.
Estimates of the accretion rate resulting from fallback-related spherical accretion could indeed help to drive a low-power outflow and contribute to the total luminosity of these sources.
\keywords{ISM: planetary nebulae: general   -- ISM: planetary nebulae: individual Hb\,12, Hen\,2-104,  MyCn\,18   --   
ISM: jets and outflows -- Stars: evolution}
}

   \authorrunning{W.A. Baan, H. Imai \& G. Orosz}            %author_head in even pages
      \titlerunning{Fallback in bipolar pre-PN }  % title_head in odd pages

   \maketitle

\section{Introduction}
\label{sec:introduction}

Planetary Nebulae (PNe) represent the post-evolution stages of medium mass stars leaving the 
Asymptotic Giant Branch (AGB) before its stellar core eventually turns into a White Dwarf (WD) \citep{Balick2002}. 
At the end of AGB evolution a superwind depletes the AGB envelope until the stellar structure changes, the 
photospheric radius shrinks, and the stellar effective temperature rises \citep{deMarco2009}.
The further evolution of PNe is then affected by the occurrence of multiple `late and very-late thermal pulse cycles' at 
the stellar core \citep{Balick2002}, the fallback of AGB shells \citep{PerinottoEA2004,ChenFBN2016}, and the ionisation 
of surrounding material by the central star, which make the PN move back and forth across the Hertzsprung-Russell 
(HR) diagram \citep{WernerH2006}.
Radiation hydrodynamical modelling of AGB-to-PN evolution shows how the central-star mass and the (post-)AGB mass-loss 
history determine the shape and the kinematics of the resulting planetary nebula \citep{PerinottoEA2004, SchonbernerEA2005}. 

The complexity of the evolutionary processes of PNe can provide for the wide variety in colour and shape of spherical 
and elliptical PNe. However,  the existence of spherical shells and ionization fronts does not explain any of the structures 
observed in bipolar PN \citep{Balick2002}. 
The small number of bipolar (elongated) PN structures indicate the existence of a directed outflow that has been described as a 
stellar wind emanating from the inner layers of the stellar structure and blowing out along the symmetry axis \citep{MellemaF1995,Balick2002}.
A few sources exhibit a well-defined `hourglass' axi-symmetric structure that defines the interaction boundaries 
of an expanding outflow with the surrounding circumstellar material \citep{SahaiEA1999}. 
Alternatively, bipolar sources have been associated with a companion star in order to create a system asymmetry 
or have been explained by accretion in a symbiotic binary system (SySt), as in the case of the Twin Jet Nebula 
\citep{CorradiEA2011}. 
One out of five PNe have indeed been confirmed to host a binary companion \citep{MiszalskiMM2015} but the evidence 
for a `binary hypothesis' is not yet very strong as compared with the `single star hypothesis' for PNe \citep{deMarco2009}.

Two of the prominent hourglass sources, Hb\,12 and MyCn\,18,  show a Spiderweb arc structure in the 
equatorial plane of the system (perpendicular to the hourglass) that has (following tradition) been explained by 
repeated stellar wind outflows  \citep{ClarkEA2014,SahaiEA1999}, by sequential ejections resulting from a thermal 
nuclear runaway \citep{KwokHsia2007}, or by episodes of activity during earlier evolutionary stages \citep{VaytetEA2009}. 
However, there is no clear explanation for a simultaneous outflow along the polar axis and the 
structure in the equatorial plane.
Similarly, intermittent jets or winds blown by an accreting binary system have also been invoked to explain 
both the inner arc structure of the Red Rectangle (HD\,44179) and its (perpendicular) outer structure \citep{Soker2005,BujarrabalEA2016}. 
Since the bipolar PNe do not show any ionised shells that are characteristic of evolved PN, 
they seem to be either at a very early evolutionary stage or at the end of PN evolution.
The suggestion has indeed been made for these to be low-excitation and young sources that only recently started 
ionising their surrounding shells and represent a short transition stage toward full development \citep{ClarkEA2014}.

The evolution of PN is known to be strongly affected by fallback by material deposited in the surrounding interstellar 
environment by AGB and PN evolutionary activity. 
However, no clear scenario has been established to recognise when fallback actually occurs. 
While fallback may happen unnoticed observationally when the accretion rate is very low and streams of fallback material 
may leak into the atmosphere of the central star of the PN, a definite accretion signature should be established when 
the accretion rate become high.
When the accretion rate becomes higher because of the availability of accretion material,  sub-Eddington accretion may happen where the low luminosity of the PN source cannot stagnate the accretion flow. 
The power generated during this (possibly short-lived) accretion could still be sufficient to energise  activity in the PN.
Although existing explanations for bipolar PN sources have traditionally relied on nuclear activity of the 
source itself, an accretion scenario should also be considered to drive an outflow along the polar axis and also explain any 
structures close to the equatorial plane.

In this paper, we seek qualitative evidence from the published literature that would support the possibility of semi-spherical 
sub-Eddington accretion unto a bipolar source such as PN Hb\,12. 
The origin of the accreting material may be the ejecta during earlier AGB stages that have been deposited in the surrounding 
medium of a post-AGB environment of a bipolar PN. 
Alternatively, accretion could also occur during or at the end of the PN evolutionary path with diminished/ceased stellar activity but 
the accretion rate from the ISM may not be sufficient during these stages.
The evidence for an accretion scenario in a bipolar source such as PN Hb\,12 will set the stage for further research 
of bipolar PN.
After describing the structural evidence of bipolar PN Hb\,12 in Section 2,  available observational evidence 
is presented in Section 3 that would directly support or be consistent with an accretion-outflow scenario.

%%%%%%%%%%%%%%%%%%%%%%%%%%%FIG %%%%%%%%%%%%%%%%%%%%%%%%%%
\begin{figure}
\begin{center}
% normal
\includegraphics[width=0.484\columnwidth,angle=0]{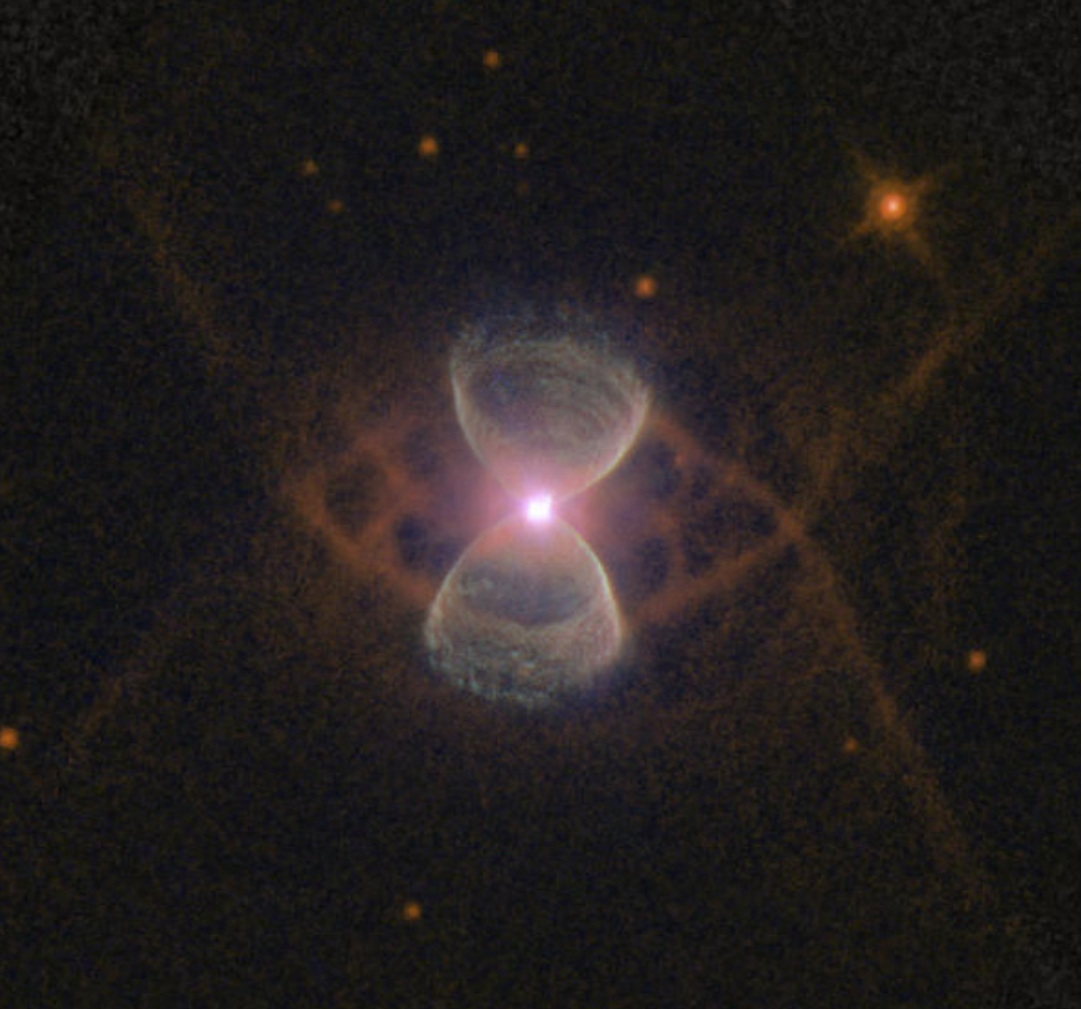}
\includegraphics[width=0.45\columnwidth,angle=0]{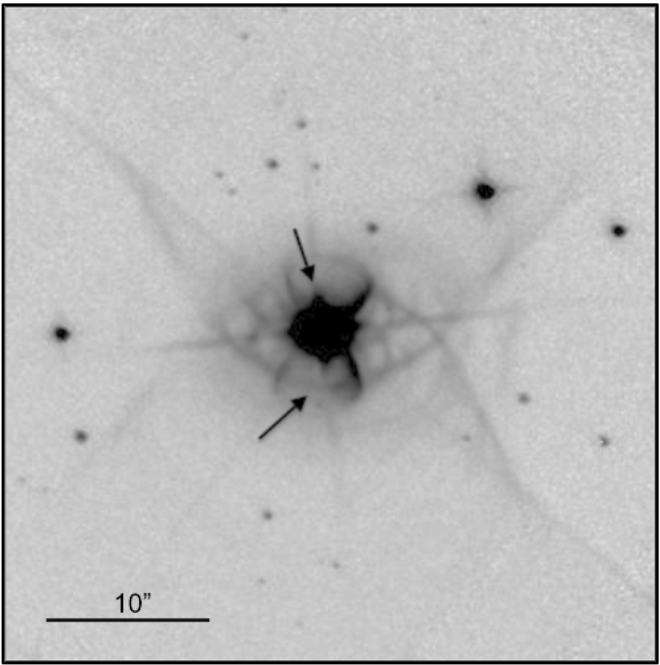}
\caption{(Left) A composite optical/infrared HST image shows the inner structure of the Matryoshka Nebula Hubble Hb\,12
with the [N\,II]($\lambda$6584 $\mu$m) emission and the NICMOS F160W image showing shocked 
[Fe\,II]($\lambda$1.6435 $\mu$m) emission in the boundary of the bipolar Outflow. Photo: ESA/NASA/J. Schmidt.
(Right) HST NICMOS F160W image from the HST Legacy Archive showing the detailed [Fe\,II] emission structures of the Spiderweb. 
The arrows point to the Spiderweb arcs detected in NICMOS F220W H$_2$ ($\lambda$2.1214 $\mu$m) data. 
The equatorial arcs are integrated with an outer and larger bipolar Hourglass structure. The field of view is 40$\arcsec$x40$\arcsec$. 
Image from \citep{ClarkEA2014}.\label{fig:spiderweb}}
\label{fig1}
\end{center}
\end{figure}
%%%%%%%%%%%%%%%%%%%%%%%%%%%/FIG %%%%%%%%%%%%%%%%%%%%%%%%%

%%%%%%%%%%%%%%%%%%%%%%%%%%%FIG %%%%%%%%%%%%%%%%%%%%%%%%%%
\begin{figure}
\begin{center}
% normal
\includegraphics[width=0.6\columnwidth,angle=0]{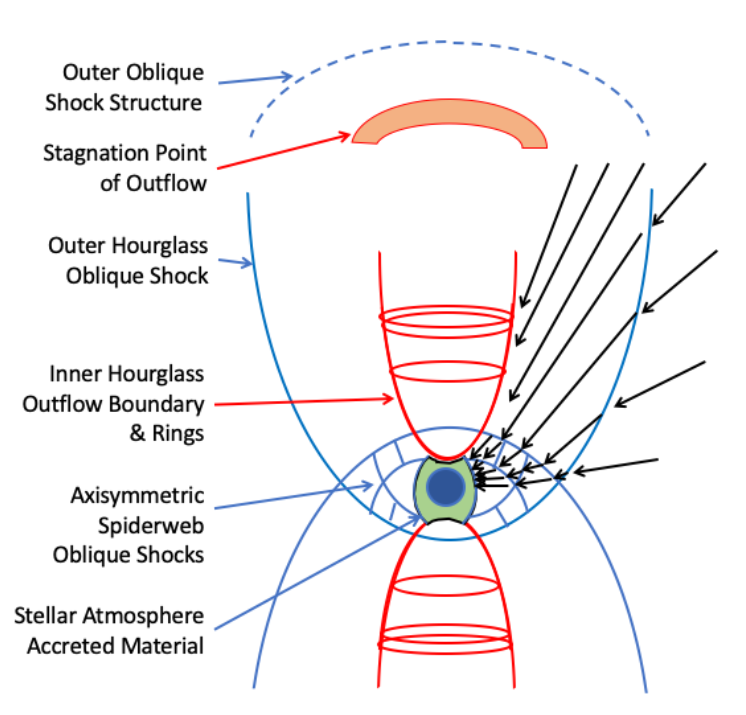}
\caption{Schematic of the structural components of PN Hb\,12. 
The red components relate to the outflow patterns and the blue components relate to inflow patterns.
The Inner Hourglass outflow boundary, the Outer Hourglass and Spiderweb oblique shocks 
are clearly seen in the images of Hb\,12 in Figure \ref{fig1}. 
The black arrows in the upper right quadrant designate the accretion flow pattern through these (blue) oblique shock structures.
A stagnation point of the (red) outflow in Hb\,12 has also been observed at a significant distance of 75$\arcsec$ from the star. 
\label{fig2}}
\end{center}
\end{figure}
%%%%%%%%%%%%%%%%%%%%%%%%%%%/FIG %%%%%%%%%%%%%%%%%%%%%%%%

%\vspace*{5mm}
\section{The bipolar structure of PN Hb\,12} \label{sec2}

The bipolar Matryoshka Nebula PN Hb\,12 has been very well studied and provides specific fundamental insights into
 the workings of a bipolar PN \citep{HoraLatter1996, WelchEA1999, KwokHsia2007, 
VaytetEA2009, ClarkEA2014}. 
Figure \ref{fig1}(left) presents a composite optical/infrared Hubble image of the central structure of Hb\,12,
which shows a hourglass-shaped `bipolar outflow' channels seen in narrow band [N II] ($\lambda$ 6584) observations 
\citep{Balick2003,KwokHsia2007}.
The axis of the hourglass outflow structure in PN Hb\,12 with concentric ring structures in its boundary is tipped clockwise 
by 10$\deg$ and with the upper part tilted away from the plane of the sky by 25$\deg$ \citep{VaytetEA2009}. 
In addition, prominent infrared [Fe II] ($\lambda$1.6435 $\mu$m) line emission shows the two edge section of an equatorial  
and axi-symmetric Spiderweb structure together with the inner sections of a larger scale `Outer Hourglass' 
structure \citep{ClarkEA2014}. 
Figure \ref{fig1} (right) again shows the detailed [Fe II] equatorial Spiderweb structure surrounding the 
star and the continuation towards a bipolar Outer Hourglass structure that is less collimated than the (inner) outflow 
hourglass structure.
This Outer Hourglass (shock) structure is expected become less prominent with distance and will close beyond the 
stagnation point of the bipolar outflows; this closure has not yet been observed in Hb\,12.
The nearly face-on Hb\,12 displays a very detailed east -- west symmetry of the axi-symmetric Spiderweb but also 
a north -- south symmetry at either side. 
These structural details of the Spiderweb will be essential for understanding what is happening in these regions.

The bipolar outflow, the Spiderweb structure, and the outer hourglass structures found in Hb\,12 and other bipolar 
sources are the defining structural components to be explained consistently in modelling these sources.
Prominent co-axial structures emanating from the core region are also found in the Hourglass Nebula PN MyCn\,18 
\citep{BryceEA1997}, as well as in  symbiotic bipolar nebulae such as the Southern Crab Hen\,2-104, Menzel's young 
PN Ant Nebula Mz\,3,  Minkowski's Twin Jet (or Butterfly) Nebula M\,2-9 
\citep[][and references therein]{CorradiEA2001,ClyneEA2015}, and the post-AGB pre-PN (Rotten) Egg Nebula 
OH\,231.8+04.2  \citep[see][]{BalickEA2020}.

In this paper we adopt the distance to Hb\,12 to be 2.24 kpc, which renders the 15$\arcsec$ extent of these structures to be 
~0.16 pc (34000 AU). In reality the distances to Hb\,12 (estimates from 2.24 to 14.25 kpc) and other PNe are highly uncertain, 
which makes size determinations difficult  \citep{ClarkEA2014}. 
Considering the extreme detail observed in the Spiderweb and the bipolar outflow in PN Hb\,12, the distance may actually be 
less than the 2.24 kpc estimate. 
Further hydrodynamic modelling of the Spiderweb structures may facilitate a more accurate distance determination.

\section{Evidence for outflow and inflow }

The presence of outflows in bipolar PN has been long recognised in the literature but 
the simultaneous occurrence of a bipolar outflow combined with Spiderweb and Outer Hourglass structures have not 
led to satisfactory explanations.
In the following sections, we consider the existing literature for any evidence that accretion inflow may be driving the outflows in 
bipolar sources and that an accretion scenario consistently explains the nature of all observed structural components in Hb\,12.

\subsection{A Semi-Spherical Accretion and Outflow Scenario} \label{sec3.1}

The outflow structure, the Spiderweb, and the outer structures observed in Hb\,12 are delineated by forbidden [N II] 
($\lambda$6584) and infrared [Fe II] ($\lambda$1.6435 $\mu$m) line emissions that all may be assumed to be 
related to shock heating and excitation.
While the line emission associated with the outflow may be easily understood, the other line emission structures 
would suggest a non-outflow explanation.
The existence of an accretion inflow close to the stellar surface would naturally result in an axisymmetric system of oblique 
shocks that guides the initially radial accretion inflow into the equatorial region and around the outflow structure as captured in the 
diagram of Figure \ref{fig2}). 
Hydrodynamically an oblique shock stands under an angle with the flow direction and will thermalise part of the flow 
velocity component perpendicular to the shock, while leaving the parallel velocity component intact.
Assuming that the underlying star has ceased its own shell burning activity and has no 
substantial stellar wind, the oblique shocks of the Outer Hourglass serve to decelerate and redirect the in-falling gas at large 
distances in order to guide the flow around the bipolar outflow regions and into the axisymmetric oblique shocks of the 
Spiderweb structure in the equatorial region. 
The very structured shock patterns observed around PN Hb\,12 and other sources may thus be consistent with a 
hydrodynamic inflow pattern resulting from accretion, either spherical or from a distant companion star.
While full hydrodynamical modelling of this inflow-outflow structure will need to further confirm this scenario, it will also provide a measure of the accretion rate, the strength of the oblique shocks, and the actual scale size of the shock patterns (and indirectly 
the distance).

\subsection{Bipolar Outflow in the Inner Hourglass} \label{sec3.2}

The bipolar Inner Hourglass outflow structure in Hb\,12 represents a prominent feature in the source with the base seen 
in forbidden [Fe II] line emission and the outline of the whole structure seen in the lower density tracing forbidden 
[N II] emission (Fig. \ref{fig1}).
The observed structure clearly suggests a low power outflow expanding into a pressure gradient and energising a 
widening boundary layer as outlined by the [Fe II] and [N II] line emission \citep{KwokHsia2007,VaytetEA2009}.
The shape of the boundary of this hourglass outflow channel 
suggests consistency with a supersonic and accelerating fluid outflow from a simple DeLaval-like 
nozzle expanding into a circumstellar medium with a radial pressure profile p(r) $\propto$ r$^{-2}$ \citep[using the 
formalism from][]{LandauL1959, Choudhuri1998}. 
The boundaries of this outflow are self-adjusting and are determined by the equilibrium between 
the internal pressure of the accelerating flow and the radially varying ambient pressure in the circumstellar envelope. 
The supersonic outflow along the polar axis starts at a narrowed throat formed in the atmosphere where the gas 
reaches a proton escape velocity equaling the local velocity of sound. 
If the 90 \kms base velocity is indicative of the local escape velocity from a 1 \msol \ star, then this Mach nozzle may 
form at a radius of about 10$^{12}$ cm and require a local temperature of 10$^5$ K, which is higher than the effective 
surface temperature of 35,000 K \citep{HyungA96}. 
Our first order modelling of the case with a quadratic pressure gradient shows that the flow accelerates early in the nozzle 
and may reach a final velocity of at least 2.5 times the sound velocity at the throat of the nozzle. 
Sustained operation of the nozzle outflow would require sustained replenishment of heated atmospheric material.
Detailed modelling may provide an estimate of the power and the scalesize of the outflow.

The boundary layer observed in Hb\,12 and other sources will naturally develop around the outflow where the edges 
of the flow pattern are slowly decelerating because of energy and momentum being dissipated. 
This boundary layer of Hb\,12 is very prominent with shock-tracer [Fe II] emission lines at the onset of the outflow 
and is seen in [N II] emission away from the star as the interaction becomes less intense  \citep{ClarkEA2014}.
The velocities at the base of the Hb\,12 outflow are estimated to be 90 $\pm$15 \kms in the redshifted (north) and blueshifted (south) 
asenses and consistent with the velocity estimate mentioned above.
Similar bipolar structures and outflow velocities have been detected in the PNe MyCn\,18 (V = $\pm$48 \kmss) 
\citep{CorradiS1993}, in Mz\,3 (V = 130 \kmss) \citep{SantanderEA2004}, and the Inner Hourglass structure of 
the symbiotic Mira Hen\,2-104 (V = 50 \kmss) \citep{CorradiEA2001,ClyneEA2015}.

At a certain distance from the central star, bipolar outflows are expected to stagnate and produce lobes containing
accumulated outflow and entrained material that are analogous to the radio lobes of extragalactic sources.
The stagnation lobes in the low-power outflow of PN Hb\,12 appear as multiple knots at a distance of 75$\arcsec$ (0.84 pc if 
at 2.24 kpc) that are emitting low-ionisation [N II] line emission \citep{VaytetEA2009}. 
The velocities of these knots are at $+$275 (north)  and $-$296 \kmss, which is slightly higher than a few times the 
 base velocity 90 \kmss.
Stagnation points have also been identified in the sources Hen\,2-104 and the Rotten Egg Nebula
 \citep{CorradiEA2001,KwokHsia2007}. 

The multiple coaxial rings in the walls of the hourglass outflow structures of Hb\,12 are also seen as low-ionisation 
[N II] thermal emission \citep{KwokHsia2007} and would result from energy dissipation in the boundary layer (see Fig. \ref{fig1}).  
These rings appear roughly equally spaced at approximately 850 AU (if at 2.24 kpc) and may result from some 
axi-symmetric Kelvin-Helmholtz-like instability resulting from the velocity gradients across the boundary layer.
Alternatively, these rings could represent decelerating shocks only in the boundary layer as a result 
of the decreasing ambient pressure shaping the outflow.
Similar rings are also prominent in the sources MyCn\,18 \citep{SahaiEA1999}, M\,2-9 \citep{KwokHsia2007}, 
and Hen\,2-104 \citep{CorradiEA2001}.

\subsection{Evidence for inflow at the Outer Hourglass}\label{sec3.3}

The Outer Hourglass oblique shock structure appears as two axisymmetric elliptical surfaces that reach around
the star and interconnects with Inner Spiderweb at its outer edges (see Fig. \ref{fig2}).  
In PN Hb\,12 only the base of the shock-related [Fe II] Outer Hourglass emission structures can be seen in 
(Fig. \ref{fig1}) but they are expected to extend towards the stagnation points in the outflows \citep{VaytetEA2009}. 
The axisymmetric region between the boundary layer of the Inner Hourglass outflow and the Outer Hourglass shocks
would also be filled with in-falling material that has been shock-heated and may show H$_2$ and [N II] tracer emission.
All shock-related [N II] and H$_2$ emissions within the Outer Hourglass structure are expected to decrease in 
intensity away from the star because of decreasing densities and decreasing perpendicular velocities in the oblique shocks.
While Hb\,12 does not (yet) exhibit such emissions, the filamentary [N II] $\lambda$6583 emitting cloudlets in 
PN Hen\,2-104 trace a very similar Outer Hourglass structure far beyond the Inner Hourglass outflow 
\citep{CorradiEA2001,Corradi2003,KwokHsia2007}, 
Similar emissions are found in the Twin Jet Nebula M\,2-9, and PN Mz\,3 \citep[see][]{KwokHsia2007, ClyneEA2015}.  

A long-slit velocity-position diagram taken along the major axis of the Hen\,2-104 system shows that the bipolar 
outflow has (accelerated) positive velocities in the north up to projected $+$90 \kmss, and negative velocities 
in the south \citep{CorradiS1993,CorradiEA2001}. 
However, the diagram also shows opposite velocities in the north as well as in the south that increase up to projected 
of 30 \kmss at the edge of the Outer Hourglass structure.
While the bipolar Inner Hourglass shows an outflow pattern, the filaments inside the Outer Hourglass  of Hen\,2-104 
would indeed represent the expected inflowing accretion component.

\subsection{Inflow through the Spiderweb Structures} \label{sec3.4}

The apparent shock structures in both the equatorial Spiderweb and the Outer Hourglass structures are consistent with 
a large scale hydrodynamic flow pattern leading towards the stellar core in Hb\,12.
The series of oblique shocks redirect and decelerate the accretion flow, 
raise the density and the temperature, and funnel the flow into the equatorial region.
The shock nature of these structures may be supported by the detection of the shock tracer [Fe II] ($\lambda$1.6436 $\mu$m) 
emission, continuum ($\lambda$1.6$\mu$m), and the H$_2$ ($\lambda$2.1214 $\mu$m) line emission  \citep{HoraLatter1996,ClarkEA2014}.  
The two tangential edge sections of this axisymmetric system of standing oblique 
shocks show in great detail how the inflow is being guided onto an equatorial 
surface of the star (Figure \ref{fig2}).
The nearly face-on position of Hb\,12 also shows the north-south symmetry in the Spiderweb, which indicates that 
the flow pattern near the star is the same above and below the equatorial plane.
Incidentally, these oblique shocks work on the same principles as a diffuser (air intake) of a supersonic airplane.
The inner Spiderweb and the Outer Hourglass form an interconnected oblique shock network that takes the initially 
spherical accretion flow, bypasses the region of the bipolar outflow, and brings it into the equatorial region of the star. 

Analysis of the molecular hydrogen emission in Hb\,12 suggests that collisional processes would dominate over UV 
absorption for exciting the H$_2$ emission when the kinetic temperature T$_k$ $>$ 1000 K \citep{HoraLatter1996}.
The temperature of free-falling material at a distance R from a star would be considerably high at T$_{ff}$ = 
1.08 x 10$^5$ M$_\odot$ R$_{AU}^{-1}$ K. This suggests that shock heating of the accreting material would completely 
dominate the excitation of H$_2$ inside a radius of about 100 AU from the star.  

In addition to PN Hb\,12, evidence for similar Spiderweb structures has been found in bipolar PNe such as MyCn\,18 
\citep{BryceEA1997}, Henize 2-104 \citep{CorradiEA2001,ClyneEA2015}, R\,Aqr \citep{CorradiS1995}, and Mz\,3 
and M\,2-9 \citep{ClyneEA2015}. 

\subsection{Infall in the vicinity of the stellar surface}\label{sec3.5}

A final standing shock must exist above the equatorial stellar surface in order to further decelerate the (then) nearly 
radial flow pattern to a subsonic velocity and to let this inflow integrate with the stellar atmosphere and a possible
circumstellar disk.
Existing observational data of the H$_2$ and [Fe II] emissions in Hb\,12 taken just north of the core, at the core, 
and just south of the core provide the evidence for this inflow inside the Spiderweb of Hb\,12 \citep[Figure 2 in][]{ClarkEA2014}.
At the `northern edge' of the nuclear region and covering the edge of the Inner Hourglass, the shock 
tracer [Fe II] emission shows a redshifted northern outflow with a (projected) velocity in the range of $+$30 to $+$60 \kmss. 
However, there are also positive velocities in the [Fe II] and H$_2$ data south of the core at the edge of the Spiderweb 
in the range 0 to $+$30 \kmss.
Similarly, the blueshifted `southern' outflow is seen south of the core in the [Fe II] data at $-$60 to $-$30 \kmss, while the  
H$_2$ data also shows emission north of the core at the Spiderweb structure in the range  $-$30 to 0 \kmss.
While the strongly positive and negative velocity components in the [Fe II] data can be explained as being part of the 
bipolar outflow, the smaller velocity components in the same direction at the opposite sides of the core do not fit the picture.
A simple interpretation of these components would be that velocities in the same direction (positive or negative) at either 
(north and south) side of the tilted source indicate that one can represent an outflow and then the other is an inflow.
This suggests that the observational data presented by \cite{ClarkEA2014} show a outflow-inflow velocity pattern that is 
consistent with an accretion scenario. 

\subsection{Evidence for a disk around the central star} \label{sec3.6}

The accretion flow into an equatorial footprint may result in the formation of a thick disk/torus 
around the central star as a form of temporary mass storage before material transfers into the atmosphere of the star.
Evidence for such an equatorial torus has indeed been found in Hb\,12  using the bright He I (2.0585 $\mu$m) and Br$\gamma$ 
(2.1655 $\mu$m) recombination line emission showing a central cavity with a 0.2$\arcsec$ x 0.41$\arcsec$ (450 $\times$ 
930 AU if at 2.24 kpc) 
inner torus structure  \citep{ClarkEA2014}.
However, the observed velocity profile perpendicular to the symmetry axis indicates a rotation velocity of 90 \kmss, which 
would suggest an orbital radius in the disk of only 0.22 AU around a one \msol  \ star (and an uncomfortable distance 
of only 2.2 pc to the observer).
  
The presence of a disk in a bipolar PNe may also be confirmed by any maser activity to be detected at their central regions.
OH (1665 MHz) and H$_2$O (22.12 GHz) maser features at the core region of the bipolar PN K\,3-35 suggest 
a disk with an extent of 98 AU (if at 5 kpc) \citep{MirandaEA2001}. 
In addition,  the OH and H$_2$O features at the core K\,3-35 are redshifted relative to the OH (1667 MHz) maser components 
found in the bipolar outflows, which would further confirm inflow into the central region.  
OH and H$_2$O masers are also found in the core regions of other bipolar PNe \citep{GomezEA2008, 
UscangaEA2014, GomezEA2015}.
Similarly, the in-falling SiO (43 GHz) masers in the bipolar post-AGB pre-PN (Rotten Egg) OH\,231.8+04.2 
have been interpreted as being part of a torus  \citep{SanchezEA2002}.

The shocked-heated material in the disc around the stellar core will contribute to emission of the source in addition to 
ongoing or enhanced shell burning possibly triggered by the accreted material. 
The estimated effective surface temperature of Hb\,12 is 35,000 K based on modelling of the observed emission lines 
\citep{HyungA96}, which suggests an early O-star classification and would indicate an early stage in 
PNe evolutionary models \citep{PerinottoEA2004,WernerH2006}.

\subsection{The possibility of sub-Eddington Accretion} \label{sec3.7}

Considering that bipolar PNe do not (yet) exhibit the shell-like characteristics of spherical and elliptical PN, the 
possibility should be considered that the bipolar activity is driven by accretion in sources such as Hb\,12. 
For early-type PN the material may be accreted from a circumstellar cocoon/envelope or a large circumbinary disk that was 
formed during the later stages of post-AGB evolution \citep{PerinottoEA2004,Herwig2005,AkashiS2008}. 
Simulations show that depending on the mass of the underlying star, such envelopes/shells may exist at 
distances ranging from a few to 10 $\times$ 10$^{17}$ cm and have particle densities ranging from 20 to a few 
hundred cm$^{-3}$ \citep{PerinottoEA2004,ChenFBN2016}.
During the early evolutionary PN phases lasting a few 10$^4$ yr, the free-fall timescale 
$t_{ff} = R^{3/2} / (2GM)^{-1/2}$ suggests that material within a radius of 
$R$ = 0.22 $\times$ 10$^{17}$ cm can fall back unto the star.

An accretion model that most adequately describes sub-Eddington accretion from a (distant) surrounding medium is 
the Bondi-Hoyle spherical accretion model  \citep{Bondi52,Mestel54}. 
While the luminosity of the source is not sufficient to halt the spherical accretion, a hemispherical balance may be 
established between the polar outflows and equatorial accretion flow regions, while the accreting 
material away from the star does not yet know about any complicated flow pattern close to the star.
Assuming a uniform surrounding medium, a spherical accretion rate may be expressed for the temperature limited 
case with $\gamma$ = 3/2 as \citep{Bondi52}:

${\dot M_{acc}} = 2 \pi \alpha (G M_{st}^2 (V^2 + c_s^2)^{-3/2} \rho_{ism}) = 
 3.5 \times 10^{-9} M_{st}^2\, n_{100}\, T^{-3/2}_{10}\, \msol/yr, $
\label{eq1}

\noindent where $M_{st}$ is the stellar mass assumed to be 1 \msol , $V$ is the spatial velocity of the star 
and is assumed to be zero, $c_s$ is the speed of sound that depends on the temperature of the surrounding medium 
expressed in units of $T_{10}$ = 10K, and $\rho_{ism}$ is the particle mass density of the medium far from the star expressed 
in units of $n_{100}$ = 100 cm$^{-3}$.  
The factor $\alpha$ is between 1 and 2 depending on environmental conditions and a value of 1.5 is adopted here.
The accretion rate favours higher stellar mass, higher gas densities, and  lower temperatures. 
The luminosity resulting from the estimated accretion rate may be estimated as:

$L_{acc} =  \dot M_{acc} G M_{st} /R_{st} =  2.9 \times 10^{33} M_{1}^{3}\, n_{100}\, T^{-3/2}_{10}\,R_{10}^{-1}\, erg/s,$

\noindent where the stellar radius $R_{st}$ is expressed in units of 10$^{10}$ cm.

An estimate of the luminosity of Hb\,12 may follow from the estimated effective surface temperature of 35,000 K 
based on modelling of the observed emission lines \citep{HyungA96}, which results in a (minimum) blackbody 
luminosity for a stellar core with an effective radius $10^{11}$ cm to be $L_{bb} = 8.8 \times 10^{36}$ erg/sec.
This value is less than the Eddington Luminosity L$_{Edd}$ = 1.2 $\times 10^{38} M_{1}$ erg/sec, which would 
verify that sub-Eddington accretion is possible for PN Hb\,12.  
Assuming some reasonable parameters for the surrounding environment, a scenario with spherical accretion from 
fallback of circumstellar material could account for part of the apparent luminosity of the star and result in 
driving the outflow activity of Hb\,12. 

\section{Conclusions} \label{sec4}

The  presence of bipolar outflows seen in a subset of the PNe population may naturally result from  
accretion unto the star and the existing observational evidence of bipolar PN Hb\,12 and similar sources would (already) 
support such a scenario.
Published literature provides the evidence that fallback of gaseous shells deposited during the earlier stages of post-AGB 
and pre-PN evolution may indeed provide sufficient material for sub-Eddington accretion.
The shape of the boundaries of the low-power outflow in Hb\,12 suggests 
that it is an accelerating and supersonic flow into an environment with a quadratic pressure gradient. 
A DeLaval-like nozzle may form in the stellar atmosphere where the gas reaches escape velocity equalling the sound 
speed from where the flow expands into the pressure gradient.
The supplement of the accretion energy to the stellar atmosphere would be a trigger for establishing a bipolar outflow. 
The low luminosity of the stellar PN core is unable halt the accretion flow and establish an omni-directional stellar wind.

The unusual Spiderweb and Outer Hourglass structures observed in PN Hb\,12 (and in similar sources) may be interpreted 
hydrodynamically as oblique shocks that decelerate the initially spherically accretion flow and 
redirect the flow pattern around the hemispherical sections of the bipolar outflows and into the equatorial region of the star. 
Evidence for infall follows from a detailed inspection of the imaging data of the  [Fe II] and H$_2$ line emission 
of HB\,12, where inflow velocity components are indeed co-located with outflow velocities in the core region. 
Similarly, the axi-symmetric region between the Inner Hourglass boundary layer of the bipolar outflow and the oblique shocks 
of the Outer Hourglass would be filled with inflowing gas, as seen in emission in the similar Spiderweb source Hen\,2-104. 
Also the presence of an accretion disk surrounding Hb\,12 with a clear velocity gradient confirms that infalling material is collecting around the star. 
The estimate of the mass accretion rate using the Bondi-Hoyle model for spherical accretion shows that for reasonable 
parameters for the surrounding ISM environment, accretion of fallback shells could add significantly to the luminosity of 
the pre-PN source and provide enough energy to drive the bipolar outflows. 

An episode with sub-Eddington accretion undoubtedly defines a separate evolutionary stage in the life of a PN.
If the accreting material results from post-AGB activity, the star will acquire new material in its atmosphere, although some will be bipolar-ejected again and may fall back at a later time. 
After this accretion stage is finished, the star will initiate a further path of PN evolution with a new atmosphere. 
If this type of accretion happens during a later evolutionary stage of the PN, this will mark a temporary interruption of the 
PN activity to be followed by renewed PN activity with a replenished atmosphere.
Since the availability of material to be accreted is an important factor, episodes of accretion depend very much on the details
of AGB to PN and it is likely that stages of accretion will occupy only part of the 
evolutionary lifetime of a PN.

Further verification of the hydrodynamic interpretation of the accretion scenario will follow from (ongoing) modelling of the spherical 
Bondi-Hoyle accretion flow and the oblique shocks of the Spiderweb and the Outer Hourglass structures. 
Hydrodynamic modelling of the accretion flow and the supersonic DeLaval-nozzle bipolar outflow will provide an 
indication of the actual sizes of the oblique shock components and the outflow, and provide an independent estimate 
for the (uncertain) distance of these sources. 
Modelling efforts of the bipolar outflow and of the accretion flow have been initiated. 
A general verification of the accretion scenario would also result from (planned) radio observations of the shocks of the 
Spiderweb and the Outer Hourglass structure in Hb\,12 and the (potential) detection of maser action in the shocked dense 
material of the disk in the system.

%\vspace*{1mm}
\section*{Acknowledgements}
WAB thanks the Graduate School of Science and Engineering of Kagoshima University and the Department of English 
Communication of Toyo University for their hospitality during the JSPS Invitation Program for Foreign Researchers 
(S15126). 
WAB also thanks the XinJiang Astronomical Observatory for their hospitality during the Program of High-end 
Foreign Experts of the XinJiang Uygur Autonomous Region (grants 20166500004 and 20176500001) and under the 
PIFI grant No. 2019VMA0040 from the Chinese Academy of Sciences.
HI was supported by the JSPS KAKENHI program JP16H02167. 
GO acknowledges support from the ARC Discovery project DP180101061 of the Australian Government, the CAS LCWR 
Program 2018-XBQNXZ-B-021 of China, the Japanese MEXT scholarship, the Leids Kerkhoven-Bosscha Fonds 
(LKBF17.0.002) and also thanks the support and hospitality of the Joint Institute for VLBI ERIC (JIVE).

Images made with the NASA/ESA Hubble Space Telescope are from the Hubble Legacy Archive, 
a collaboration between the Space Telescope Science Institute (STScI/NASA), the Space Telescope European Coordinating 
Facility (ST-ECF/ESA), and the Canadian Astronomy Data Centre (CADC/NRC/CSA). 
%\end{acknowledgements}

\bibliographystyle{plainnat}

\begin{thebibliography}{}

\bibitem[Akashi \& Soker 2008]{AkashiS2008} Akashi, M. \& Soker, N., 2008, New Astronomy, 13, 157

\bibitem[Balick 2003]{Balick2003} Balick, B., 2003, {\it Symbiotic Stars Probing Stellar Evolution}, eds. R.L.M. Corradi et al.,  
ASP Conf. Ser., 303, 407

\bibitem[Balick \& Frank 2002]{Balick2002} Balick, B. \& Frank, A., 2002, \araa, 40, 439

\bibitem[Balick et al. 2020]{BalickEA2020} Balick, B., Frank, A., Liu, B, Huarte-Espinosa, M., 2020, \apj,  843, 108

\bibitem[Bondi 1952]{Bondi52} Bondi, H., 1952, \mnras, 112, 195

\bibitem[Bryce et al. 1997]{BryceEA1997} Bryce, M., Lopez, J.A., Hollowat, A.J., \& Meaburn, J., 1997, \apj, 478, 161

\bibitem[Bujarrabal et al. 2016]{BujarrabalEA2016} Bujarrabal, V.,  Castro-Carrizo, A., Alcolea, M, Santander-Garcia, H., 
et al., 2016, \aap, 593, A92

\bibitem[Choudhuri 1998]{Choudhuri1998} Choudhuri,  A. R., 1998,  {\it The physics of fluids and plasmas : an introduction for
astrophysicist}

\bibitem[Chen et al. 2016]{ChenFBN2016} Chen, Z., Frank, A., Blackman, E.G., \& Nordhaus, J.,  2016, \mnras, 457, 3219

\bibitem[Clark et al. 2014]{ClarkEA2014} Clark, D.M., Lopez, J.A., Edwards, M.L., et al., 2014, \apj, 148, 98

\bibitem[Clyne et al. 2015]{ClyneEA2015} Clyne, N., Akras, S., Steffen, W., Redman, M.P., et al., 2015, \aap, 582, A60

\bibitem[Corradi 2003]{Corradi2003} Corradi, R. L. M., 2003, {\it Symbiotic Stars Probing Stellar Evolution}, eds.
R. L. M. Corradi et al., ASP Conf., 303, 393

\bibitem[Corradi \& Schwarz 1993]{CorradiS1993}  Corradi, R.L.M., \& Schwarz, H.E., 1993, \aap, 268, 714

\bibitem[Corradi \& Schwarz 1995]{CorradiS1995}  Corradi, R.L.M., \& Schwarz, H.E., 1995, \aap, 293, 871

\bibitem[Corradi et al. 2001]{CorradiEA2001} Corradi, R.L.M., Livio, M., Balick, B., et al., 2001, \apj, 553, 211

\bibitem[Corradi et al. 2011]{CorradiEA2011} Corradi, R.~L.~M., Balick, B., \& Santander-Garc{\'{\i}}a, M., 2011, \aap, 529, A43 

\bibitem[de Marco 2009]{deMarco2009} De Marco, O., 2009, PASP, 121, 316

\bibitem[G\'omez et al. 2008]{GomezEA2008} G\'omez, J.F., Su\'arez, O., G\'omez, Y., et al.. 2008, \aj, 135, 2074

\bibitem[G\'omez et al. 2015]{GomezEA2015} G\'omez, J., Su\'arez, O., Benjoya, P., et al.. 2015, \apjs, 799, 186

\bibitem[Hyung \& Aller 1996]{HyungA96} Hyung, S. \& Aller, L.H.. 1996, \mnras, 278, 551

\bibitem[Herwig 2005]{Herwig2005} Herwig, F.. 2005, \araa, 43, 435

\bibitem[Hora \& Latter 1996]{HoraLatter1996} Hora, J.L., \& Latter, W.B.. 1996, \apj, 461, 288

\bibitem[Kwok \& Hsia 2007]{KwokHsia2007}  Kwok, S., \& Hsia C.H.. 2007, \apj, 660, 341

\bibitem[Landau \& Lifshits 1959]{LandauL1959} Landau, L.D. \& Lifshitz, E.M.,1959, {\it Pergamon Press, London}

\bibitem[Mellema \& Frank 1995]{MellemaF1995} Mellema, G. \& Frank, A.,  1995, \mnras, 273, 401

\bibitem[Mestel 1954]{Mestel54} Mestel, L.,1954, \mnras, 114, 437

\bibitem[Miranda et al. 2001]{MirandaEA2001} Miranda, L.F., G\'omez, Y., Anglada, G., \& Torrelles, J.M., 2001, \nat, 114, 284

\bibitem[Miszalski et al. 2015]{MiszalskiMM2015} Miszalski, B.; Manick, R.; McBride, V., 2015, {\it The Physics of Evolved Stars}, eds. E. Lagadec et al., EAS PS, 71-72, 117

\bibitem[Perinotto et al. 2004]{PerinottoEA2004} Perinotto , M., Sch\"onberner, D., Stefffen, M. , \& Calonaci, C., 2004, \aap,414, 993

\bibitem[Sahai et al. 1999]{SahaiEA1999} Sahai, R., Dayal, A., Watson, A.M., et al., 1999, A\aj 118, 468

\bibitem[S\'anchez-Contreras et al. 2002]{SanchezEA2002} S\'anchez-Contreras, C., Desmurs, J.F., Bujarrabal, V., et al.,2002, \aap, 385, 1

\bibitem[Santander-Garc\'ia et al. 2004]{SantanderEA2004} Santander-Garc\'ia,  M., Corradi, R. L. M., Balick, B., Mampaso, A., 2004, \aap, 426, 185

\bibitem[Schonberner et al. 2005]{SchonbernerEA2005} Sch\"onberner, D., Jacob, R., Steffen, M., et al.,  2005, \aap, 431, 963

\bibitem[Soker 2005]{Soker2005} Soker, N., 2005, \aj, 129, 947 

\bibitem[Uscanga et al. 2014]{UscangaEA2014} Uscanga, L., G\'omez, J.F., Miranda, L.F., et al., 2014, \mnras, 444, 217

\bibitem[Vaytet et al. 2009]{VaytetEA2009} Vaytet, N.M.H., Rushton, A.P., Lloyd, M., et al., 2009, \mnras, 398, 385

\bibitem[Welch et al. 1999]{WelchEA1999} Welch, C.A., Frank, A., Pipher, J.L., et al., 1999, \apj 522, L69

\bibitem[Werner \& Herwig 2006]{WernerH2006} Werner, K. \& Herwig, F., 2006, \pasp, 118, 183

\end{thebibliography}

\label{lastpage}

\end{document}